\documentclass[a4paper,fleqn,usenatbib]{mnras}

 \usepackage[normalem]{ulem}
\usepackage[T1]{fontenc}
\usepackage{ae,aecompl}

\usepackage[pdftex]{graphicx}

\usepackage{amsmath}	% Advanced maths commands
\usepackage{amssymb}	% Extra maths symbols
\usepackage{multirow}
\usepackage{longtable}
\usepackage{supertabular}
\usepackage{tablefootnote}
\usepackage{epstopdf}

\usepackage{natbib}
\hypersetup{draft}

\newcommand{\emm}[1]{\ensuremath{#1}} 
\newcommand{\emr}[1]{\emm{\mathrm{#1}}}
\newcommand{\chem}[1]{\ensuremath{\mathrm{#1}}} 
\newcommand{\unit}[1]{\emr{\,#1}} 
 
\newcommand{\MHz}{\unit{MHz}} 
\newcommand{\kHz}{\unit{kHz}} 
\newcommand{\radec}[6]{\emr{\alpha_{2000}=#1^{h}#2^{m}#3^{s}},
	\emr{\delta_{2000}=#4^{\circ}#5^{'}#6^{''}}} 
\newcommand{\Kkms}{\unit{K\,km\,s^{-1}}} 
\newcommand{\kms}{\unit{km\,s^{-1}}} 
\newcommand{\ps}{\unit{s^{-1}}} 
\newcommand{\K}{\unit{K}}
\newcommand{\sciexp}[2]{\emm{#1\times10^{#2}}}
\newcommand{\pscm}{\unit{cm^{-2}}}

\title[ Cyclic-C$_3$HD in IRAS 16293-2422]{A study of singly deuterated cyclopropenylidene c-C$_3$HD in protostar IRAS 16293-2422}

\author[Majumdar et al.]{
L. Majumdar$^{1,2}$\thanks{E-mail: liton.icsp@gmail.com}, P. Gratier$^{1}$, I. Andron$^{1}$, V. Wakelam$^{1}$, E. Caux$^{3,4}$
\\ 
$^{1}$Laboratoire d'astrophysique de Bordeaux, Univ. Bordeaux, CNRS, B18N, allée Geoffroy Saint-Hilaire, 33615 Pessac, France\\
$^{2}$ Indian Centre For Space Physics, 43 Chalantika, Garia Station Road, Kolkata, 700084, India\\
$^{3}$Universit\'e de Toulouse, UPS-OMP, IRAP, Toulouse, France\\
$^{4}$CNRS, IRAP, 9 Av. Colonel Roche, BP 44346, F-31028 Toulouse Cedex 4, France
}

\date{Accepted XXX. Received YYY; in original form ZZZ}

\pubyear{2015}

\begin{document}
%\label{firstpage}
\maketitle

\begin{abstract}
%Aims
%Isotopomers (i.e. isotopic isomers) are excellent tools to study the physical and chemical properties for different astronomical environments. 
Cyclic-C$_3$HD ({\it c}-C$_3$HD) is a singly deuterated isotopologue of {\it c}-C$_3$H$_2$, which is one of the most abundant and widespread molecules in our Galaxy.  
%Methods
We observed IRAS 16293-2422 in the 3 mm band with a single frequency setup using the EMIR heterodyne 3 mm receiver
of the IRAM 30m telescope.
%Results
We observed seven lines of {\it c}-C$_3$HD and three lines of {\it c}-C$_3$H$_2$. Observed abundances are compared with astrochemical 
simulations using the {\tt NAUTILUS} gas-grain chemical model. 
%Our results clearly shows that {\it c}-C$_3$HD can be used as an important supplement for studying chemistry and 
%physical conditions for cold environments. 
Assuming that the size of the protostellar envelope is 3000 AU and same excitation temperatures for both {\it c}-C$_3$H$_2$  and {\it c}-C$_3$HD, we obtain a deuterium fraction of $14_{-3}^{+4}\%$.  

\end{abstract}


%%%%%%%%%%%%%%%%%%%%%%%%%%%%%%%%%%%%%%%%%%%%%%%%%%

%%%%%%%%%%%%%%%%% BODY OF PAPER %%%%%%%%%%%%%%%%%%

\begin{figure*}

    \includegraphics[width=\textwidth]{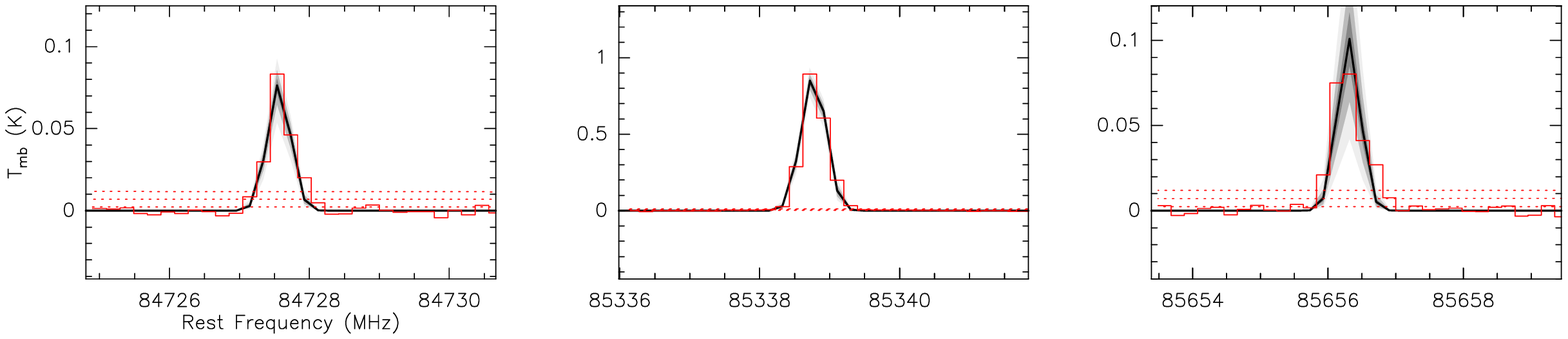}
      \includegraphics[width=\textwidth]{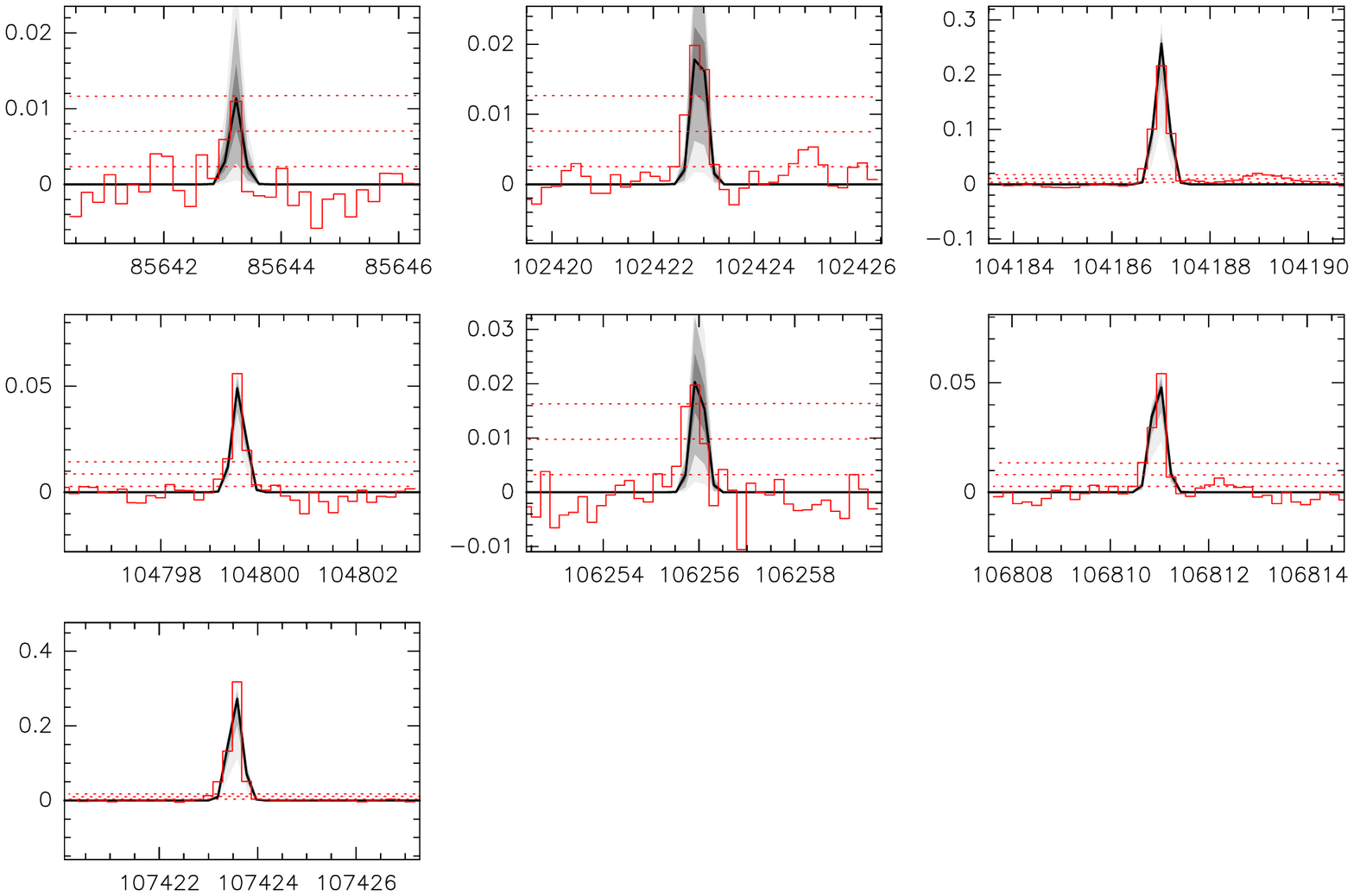}

    \caption{\label{fig.gauss_fit} Here top three and bototom seven panels are for {\it c}-C$_3$H$_2$ and {\it c}-C$_3$HD respectively. Red line: Observed lines attributed to {\it c}-C$_3$HD and {\it c}-C$_3$H$_2$. 
    Black line: Distribution of modelled spectra following the posterior distribution of parameters shown in  Figure~\ref{fig.bayes_result1} and  Figure~\ref{fig.bayes_result2}. Thick line: median 
    of the distribution. Dark grey region 68\% and respectively light grey region 95\% confidence intervals. Dotted lines are 1, 3, 5$\sigma$ noise levels.}
\end{figure*}

\section{Introduction}
The cyclic form of C$_3$H$_2$, cyclopropenylidene ({\it c}-C$_3$H$_2$), is one of the most abundant and widespread molecule in our Galaxy \citep{1985ApJ...298L..61M}. It is 
also the first hydrocarbon ring molecule found in space. It has been detected in different astronomical environments: from diffuse gas to cold dark clouds, giant molecular clouds, 
photodissociation regions, circumstellar envelopes, and planetary nebulae \citep{1985ApJ...299L..63T, 1987ApJ...314..716V, 1987A&A...181L..19C, 1989AJ.....97.1403M, 2000A&A...358.1069L}. 

 {\it c}-C$_3$H$_2$ shows strong rotational transitions in millimetre wave bands due to its large dipole moment and favourably small partition function, which make this molecule a useful probe of physical conditions 
where it is found (see \citet{1986ApJ...311L..89B} for the discussion). {\it c}-C$_3$H$_2$ is then often used as a template to study deuteration along different sources since the rotational
 transitions of its isotopic species are also very intense. {\it c}-C$_3$H$_2$ can exist in singly ({\it c}-C$_3$HD) and doubly ({\it c}-C$_3$D$_2$) deuterated forms. {\it c}-C$_3$HD was first detected in the dense 
 molecular cloud TMC-1 by \citet{1986ApJ...311L..89B}. \citet{1987A&A...173L...1G} measured a  1:5 ratio (i.e. 20\%) of the $2_{1,2} \rightarrow 1_{0,1}$  lines of {\it c}-C$_3$HD to {\it c}-C$_3$H$_2$ in TMC-1. According to 
 \citet{1987A&A...173L...1G}, this high degree of deuterium fractionation indicates that the emission comes from extremely cold dense cores (T$_{ROT}$= 4-7 K) and thus {\it c}-C$_3$HD samples the coldest part 
 of the interstellar medium (ISM). {\it c}-C$_3$D$_2$ was also detected recently by \citet{2013ApJ...769L..19S} toward the starless cores TMC-1C and L1544. They measured the abundance of {\it c}-C$_3$D$_2$ with 
 respect to the {\it c}-C$_3$H$_2$ and found it to be 0.4\%-0.8\% in TMC-1C and 1.2\%-2.1\% in L1544. This clearly shows that the abundance of {\it c}-C$_3$D$_2$ is much lower than that of {\it c}-C$_3$HD and thus supports the 
 idea that singly deuterated species forms more easily in the ISM as compared to the multiply deuterated species (\citet{1989MNRAS.240P..25B}, \citet{1998MNRAS.298..562W}). This makes {\it c}-C$_3$HD a better candidate to study 
 the chemistry and the physical conditions of cold environments compared to {\it c}-C$_3$D$_2$. The main goal of the present paper is to report the detection of {\it c}-C$_3$HD in the low mass 
 protostar IRAS 16293-2422 (hereafter IRAS 16293) and study deuterium fractionation of {\it c}-C$_3$H$_2$ from both observational and theoretical points of view.

This paper is structured as follows. In Section 2, we give a detailed description of our observations along with methodology for 
the analysis. In Section 3, we discuss the chemical model, which includes our latest deuterium chemical network with associated spin chemistry and finally results are discussed in the last Section.

\begin{table*}
   \caption{\label{tab.obs}Observed lines and spectroscopic parameters for {\it c}-C$_3$HD and  {\it c}-C$_3$H$_2$}
   \centering
   \begin{tabular}{|l| |l| |l| |l| |l| |l| |l| |l| |l| |l|}
       \hline
 Species & Lines Ref.  &  Integrated flux    &    V$_{\rm LSR}$     & FWHM  & Observed & A$_{ij}$  & E$_{up}$  & Quantum \\
       &  & (\Kkms)        & (\kms)       & (\kms) & Frequency & (\ps) & (\K) &  numbers \\
       
        &  &         &        &  & (\MHz) &  &  &   \\       
       \hline
      &                   & $0.013\pm0.009$                     & $4.3\pm0.162$                  & $1.0\pm0.70$                  & 85643.318  & \sciexp{1.43}{-5}  &  26.6   &  $4_{3,2} \rightarrow 4_{2,3}$       \\

      &                  & $0.028\pm0.003$                     & $4.3\pm0.064$                  & $1.3\pm0.12$                  & 102423.019  & \sciexp{1.53}{-5}  &  22.3   &  $4_{1,3} \rightarrow 4_{0,4}$       \\

      &                  & $0.243\pm0.024$                     & $4.2\pm0.016$                  & $1.1\pm0.02$                  & 104187.126  & \sciexp{3.96}{-5}  &  10.8   &  $3_{0,3} \rightarrow 2_{1,2}$       \\
       
   {\it c}-C$_3$HD  &\citet{1987JMoSp.122..313B} & $0.047\pm0.006$                     & $4.2\pm0.031$                  & $0.8\pm0.08$                  & 104799.707  & \sciexp{7.29}{-6}  &  10.9   &  $3_{1,3} \rightarrow 2_{1,2}$       \\
      
      &                  & $0.032\pm0.005$                     & $4.5\pm0.086$                  & $1.4\pm0.19$                  & 106256.108  & \sciexp{1.69}{-5}  &  22.5   &  $4_{2,3} \rightarrow 4_{1,4}$       \\
       
      &                 & $0.054\pm0.006$                     & $4.1\pm0.025$                  & $0.9\pm0.07$                  & 106811.09 & \sciexp{7.87}{-6}  &  10.8   &  $3_{0,3} \rightarrow 2_{0,2}$       \\
      
     &                 & $0.280\pm0.028$                     & $4.1\pm0.005$                  & $0.8\pm0.01$                  & 107423.671 & \sciexp{4.47}{-5}  &  10.9   &  $3_{1,3} \rightarrow 2_{0,2}$       \\
   \hline
   
&                  & $0.125\pm0.013$                     & $4.2\pm0.018$                  & $1.4\pm0.06$                  & 84727.688  & \sciexp{1.04}{-5}  &  16.14   &  $3_{2,2} \rightarrow 3_{1,3}$       \\
    
{\it c}-C$_3$H$_2$ &      \citet{1985ApJ...299L..63T}            & $1.372\pm0.137$                     & $4.2\pm0.010$                  & $1.4\pm0.006$                  & 85338.894  & \sciexp{2.32}{-5}  &  6.45   &   {\bf $2_{1,2} \rightarrow 1_{0,1}$ }     \\
     
&                  & $0.173\pm0.017$                     & $4.3\pm0.028$                  & $1.9\pm0.05$                  & 85656.431  & \sciexp{1.52}{-5}  &  29.07   &  $4_{3,2} \rightarrow 4_{2,3}$       \\
       
 \hline
%\multicolumn{9}{l}{\scriptsize  Notes: (a) Flux uncertainties are gaussian fit uncertainties and that a 10\% calibration error should be added.}\\
%\multicolumn{9}{l}{\scriptsize  (b) For V$_{\rm LSR}$, we have added the total uncertainties which are the quadratic sum of the gaussian fit statistical uncertainties and the frequency uncertainties.}\\
%%\tablefootnote{uncertainties and the frequency uncertainty.}
%\multicolumn{9}{l}{\scriptsize  (c) FWHM uncertainties are only the gaussian fit uncertainties.}\\
\end{tabular}
\end{table*}

\section{Observations and data reduction}

\subsection{Observations}
The observations were obtained at the IRAM 30m telescope during the period August 18-23, 2015. Overall, the 
weather condition was like an average summer (a median value of 4-6 mm water vapour). We performed our observations 
by using  the EMIR heterodyne 3 mm receiver tuned at a frequency of 89.98 GHz in the Lower Inner sideband.
This receiver was followed by a Fourier Transform Spectrometer in its 195\kHz{} resolution mode. The observed spectrum was composed of 
two regions: one from 84.4 GHz to 92.3 GHz and another one from 101.6 GHz to 107.9 GHz.

The observations were made towards the midway point between sources A and B of IRAS 16293 at \radec{16}{32}{22.75}{-24}{28}{34.2}. The 
A and B components, separated by $5.5''$, are both inside the telescope beam of our observations at all frequencies. All observations were 
performed using the wobbler switching mode with a period of 2 seconds and a throw of $90''$ ensuring mostly flat baselines
even in summer conditions and observations at low elevation. At the beginning of each observing run, the closest planet 
Saturn was used for focus. Pointing was checked every hour with mostly good pointing corrections
(less than a third of the telescope beam size $30''$).       

 \subsection{Results}
 
 \begin{table}
   \caption{Priors distribution functions for the parameters used in the
   bayesian approach.}
   \label{tab.priors}
   \centering
   \begin{tabular}{ll}
       \hline
       Parameter & Distribution  \\
       \hline
                    \emr{\log_{10} N} (\pscm)             & $\emr{Uniform}(8,22 )$ \\
                    \emr{log_{10} T_{ex}} (\K)                & $\emr{Uniform}(\log_{10} 3,\log_{10} 200)$ \\
 %                   \emr{\log_{10} S} (a.u.)              & $\emr{Uniform}(1,5)$\\
                    \emr{V} 	 (\kms)              & $\emr{Uniform}(2.8,4.8)$\\
                    \emr{\Delta V} (\kms)            & $\emr{Uniform}(0.25,10)$\\
                    \emr{\log_{10} \sigma_{add}} (\K)         & $\emr{Uniform}(-3,1)$\\
       \hline
   \end{tabular}
   \begin{minipage}{8cm}Notes:
   \emr{ Uniform}(min-value, max-value) is a uniform distribution with
   values going from minimum value to maximum value. \end{minipage}   
   
\end{table}

\subsubsection{{\it c}-C$_3$HD and {\it c}-C$_3$H$_2$ line properties}
The data were reduced and analysed using the CLASS software from the
GILDAS\footnote{\url{https://www.iram.fr/IRAMFR/GILDAS/}} package. We made 
Gaussian fits to the detected lines following a local low (0 or 1) order polynomial baseline subtraction. 
Table~\ref{tab.obs} shows the result of these fits
for the 7 observed lines of {\it c}-C$_3$HD and 3 observed lines of {\it c}-C$_3$H$_2$. In Table~\ref{tab.obs}, uncertainties on integrated flux are the Gaussian fit 
uncertainties (with an added 10\% calibration error), uncertainties on VLSR are the quadratic sum of the gaussian 
fit statistical uncertainties and the frequency uncertainties from the spectroscopic catalog, uncertainties on FWHM are the gaussian fit uncertainties only. All the observed lines are single 
component features with the mean LSR velocity of $\sim$ 4.2 km/s and the mean
FWHM of $\sim$ 1 km/s. \citet{2011A&A...532A..23C} identified four types of kinematical behaviours for different 
species based on the different FWHM and VLSR distributions. We found that {\it c}-C$_3$HD and {\it c}-C$_3$H$_2$ belong to the 
type I (i.e. FWHM$\leq$2.5 km/s, VLSR$\sim$ 4 km/s, Eup$\sim$ 0-50 K), which corresponds to species abundant in the cold envelope
of IRAS 16293. 

\subsubsection{Radiative transfer modelling for {\it c}-C$_3$HD and {\it c}-C$_3$H$_2$ }
%We use the local thermal equilibrium (LTE) radiative transfer code under GILDAS-Weeds package \citep{2011A&A...526A..47M} to 
%model the emission of {\it c}-C$_3$HD and {\it c}-C$_3$H$_2$. Figure~\ref{fig.gauss_fit} shows the comparison of observed spectra and 
%modelled spectra. Input parameters in this radiative transfer model are the species column density,
%the line width, the excitation temperature, the source size along with an accurate spectroscopic catalog. The spectroscopic catalogs for 
%{\it c}-C$_3$HD and {\it c}-C$_3$H$_2$  were retrieved from the CDMS \citep{2005JMoSt.742..215M}. All the observed frequencies along with their Einstein coefficients, 
%upper level energies and the associated quantum numbers are listed in Table~\ref{tab.obs}. 

 \citet{2016MNRAS.458.1859M} used a bayesian model to recover the distribution of parameters which
best agree with the observed line intensities for \chem{CH_3SH} in the same source. Here, we use 
the same model with the exception that not only the integrated intensities but also the full modelled spectra are compared to the observed ones. 

We use the local thermal equilibrium (LTE) radiative 
transfer code implemented in the GILDAS-Weeds package \citep{2011A&A...526A..47M} to 
model the emission of {\it c}-C$_3$HD and {\it c}-C$_3$H$_2$. The input parameters in this radiative transfer model are the species column density,
the line width, the excitation temperature, the source size along with an accurate spectroscopic catalog. The spectroscopic catalogs for 
{\it c}-C$_3$HD and {\it c}-C$_3$H$_2$  were retrieved from the CDMS \citep{2005JMoSt.742..215M}. All the observed frequencies along with their Einstein coefficients, 
upper level energies and the associated quantum numbers are listed in Table~\ref{tab.obs}. Here, we have chosen to fix the source size to $25''$ ($\sim$ 3000 AU,  typical size of the protostellar envelope  by \citet{2011A&A...532A..23C}) since the range of frequencies is small. 
The likelihood function assumes that the errors are normally distributed with a noise
 term consisting of a sum in quadrature of the observed per channel uncertainty and an additional 
noise term left as a free parameter of the model. The priors are chosen to be uniform and non informative over the range of variation as defined in Table~\ref{tab.priors}.

The sampling of the posterior distribution function is 
carried out using the Python implementation {\tt EMCEE}\footnote{\url{https://github.com/dfm/emcee}}
\citep{2013PASP..125..306F} of the Affine Invariant Ensemble Monte Carlo Markov Chain approach \citep{Goodman2010}. Sixty walkers are initialised in 
a small 6 dimensional sphere in the center of the parameter space. The chains are evolved for a burn-in sequence of 3000 steps after which the
convergence is checked by examining a plot of the running mean of each parameter.  

Figure~\ref{fig.gauss_fit} shows the comparison of observed spectra and modelled spectra. 
Figures 2 and 3 show the 1D and 2D histograms of the posterior
probability distribution function for {\it c}-C$_3$HD and {\it c}-C$_3$H$_2$ respectively.
 In both cases, all parameters are well defined. 
 %Concerning the column density and size, there is a clear degeneracy arising as shown in the corresponding 2D plots, with increasing sizes corresponding to decreasing column density until the source size 
%becomes significantly larger than the largest telescope beam.
% When looking at the marginalised distributions for the column density (topmost histogram for Figures 2 and 3), this degeneracy translates into an
% asymmetrical posterior distribution function. 
 Table~\ref{tab.bayes_result} summarises the
one point statistics for the marginalised posterior distributions of parameters. The uncertainties reported in Table~\ref{tab.bayes_result} are 1$\sigma$ symmetric error bars. 
There is about a factor 2 difference in line width between {\it c}-C$_3$HD and {\it c}-C$_3$H$_2$ (shown in Table~\ref{tab.bayes_result}), this cannot arise from opacity effects as all the lines are optically thin except {\it c}-C$_3$H$_2$ line at 85.338 GHz. 
One possibility is that {\it c}-C$_3$H$_2$ comes both from the central hotter region and the outer envelope while {\it c}-C$_3$HD only comes from the outer cold envelope. This is further strengthen by the chemical modelling which shows 
such a pattern in Figure 4. Thus, deuterium fraction we obtain in Section 2.3 is a lower limit in the external envelope.

\begin{figure*}
      \includegraphics[width=\textwidth]{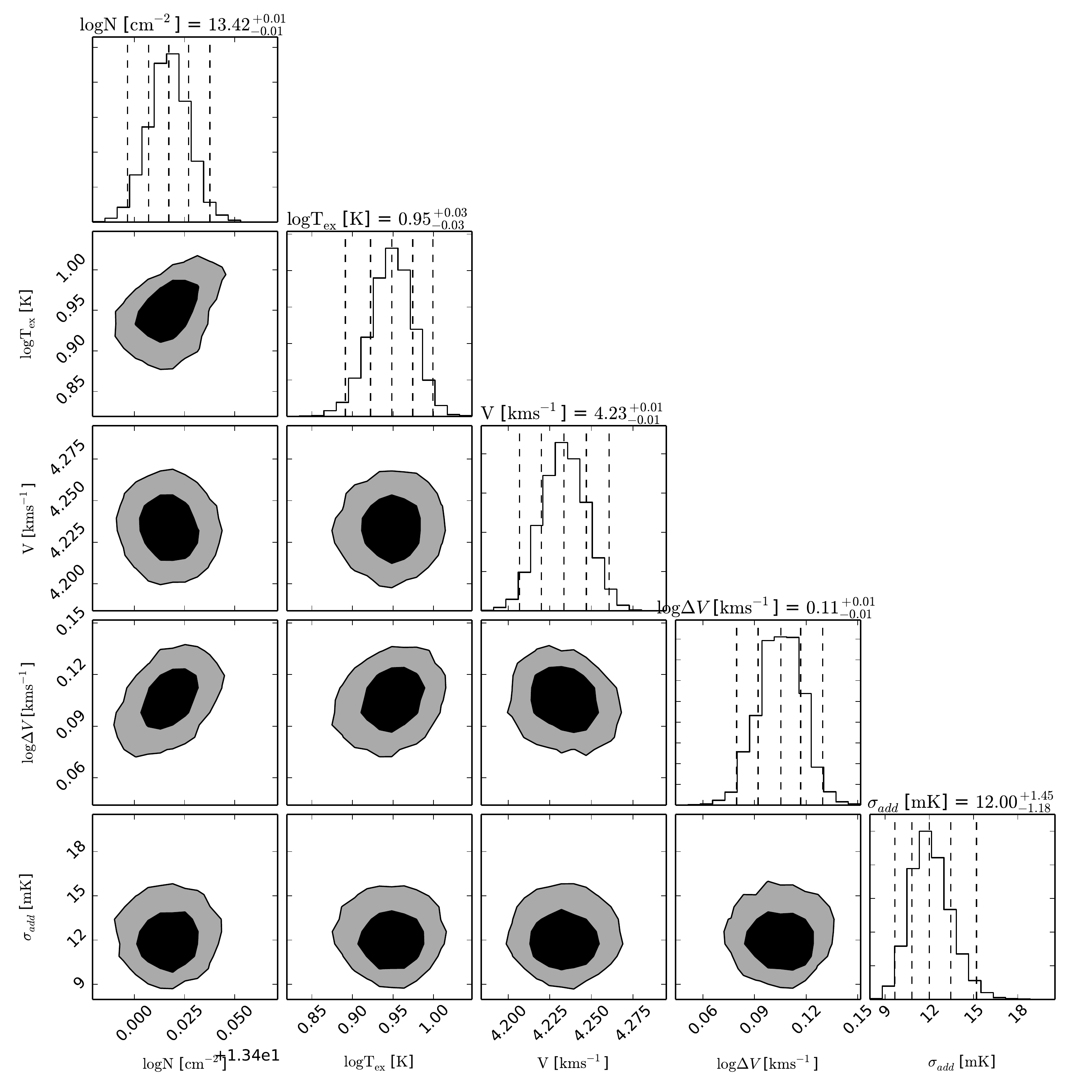}
              \caption{\label{fig.bayes_result1} 1D and 2D histograms of the
   posterior distribution of parameters for {\it c}-C$_3$H$_2$. Contours contain respectively 68 and 95 \% of samples. The quoted uncertainties are statistical only, without the 10\% calibration error.}
    \end{figure*}

    \begin{figure*}
      \includegraphics[width=\textwidth]{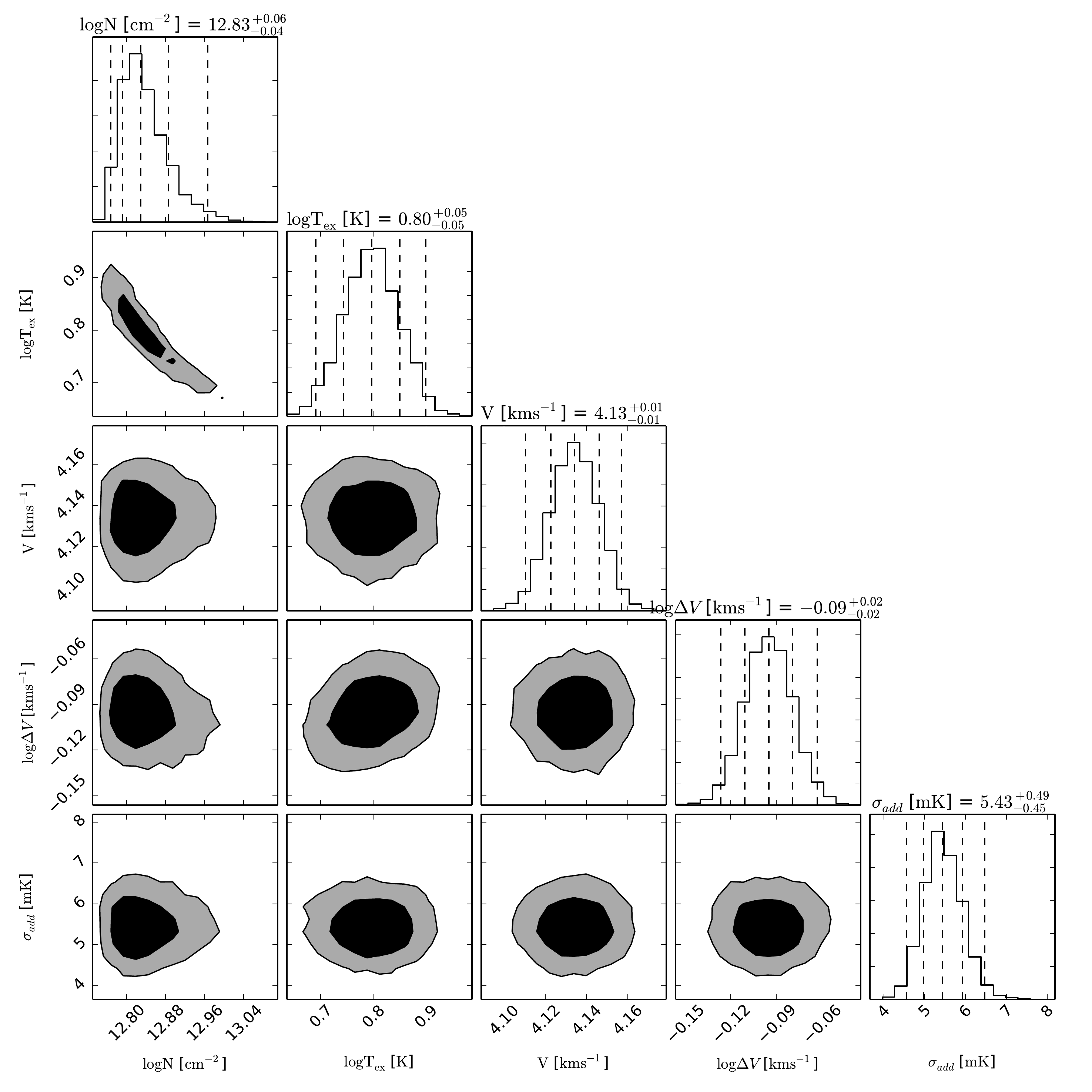}
            \caption{\label{fig.bayes_result2} 1D and 2D histograms of the
   posterior distribution of parameters for {\it c}-C$_3$HD. Contours contain respectively 68 and 95 \% of samples. The quoted uncertainties are statistical only, without the 10\% calibration error.}
    \end{figure*}

\begin{table}
\caption{Point estimates of the posterior distribution function
corresponding to the median and one sigma uncertainty.}
   \label{tab.bayes_result}
   \centering
   \begin{tabular}{lc|c}
       \hline
       Parameter      & {\it c}-C$_3$HD    & {\it c}-C$_3$H$_2$   \\
       \hline
          \emr{\log_{10} N} (\pscm)   &   $12.83\pm0.05$     &    $13.42\pm0.01$               \\
          \emr{\log_{10} T_{ex}} (\K)      &   $0.80    \pm0.05$     &      $0.95    \pm0.03$              \\
          \emr{\log_{10} \Delta V} (\kms)  &   -$0.09 \pm 0.02$  &    $0.11 \pm 0.01$             \\
  %        \emr{\log_{10} S} (a.u.)    &   $3.4\pm0.4$       &      $ \emr{\geq3.8^{b}}$               \\
          \emr{\log_{10} [X]^{a}}    &   $-11.3\pm0.05$  & $-10.7 \pm 0.01$                \\
       \hline
       \multicolumn{3}{l}{Notes: \emr{^{a}} [X] = N(X)/N(\chem{H_2}) and { \emr{^{b}} value with $95\%$ confidence.}}\\
      \multicolumn{3}{l}{Here quoted uncertainties are statistical only, without the}\\ %without the 10\% calibration error }\\
      \multicolumn{3}{l}{10\% calibration error.}\\
       \end{tabular}
\end{table}

\begin{figure*}
  \includegraphics[width=\textwidth]{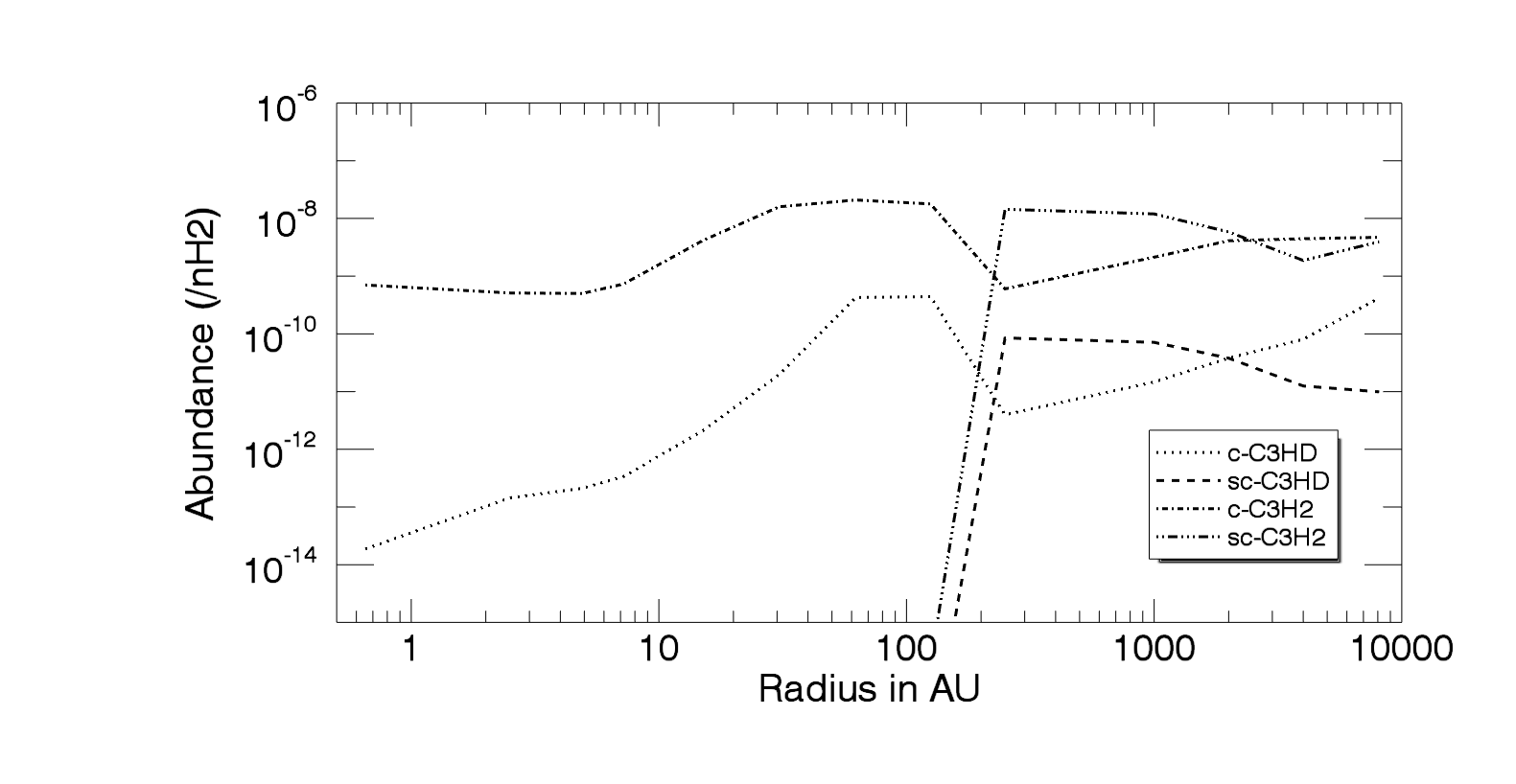}
  \caption{Abundance with respect to H$_2$ for {\it c}-C$_3$H$_2$ and {\it c}-C$_3$HD predicted by our model as a
function of radius. s{\it c}-C$_3$H$_2$ and s{\it c}-C$_3$HD represents the {\it c}-C$_3$H$_2$ and {\it c}-C$_3$HD on the surface of grains.}   
\end{figure*}

\subsection{Observational constraint on deuterium fraction for {\it c}-C$_3$H$_2$}

%\textbf{To be able to compute a deuteration fraction of {\it c}-C$_3$H$_2$, the emissions of {\it c}-C$_3$H$_2$ and {\it c}-C$_3$HD must be coming from the same region i.e. have the same angular size and excitation temperature (T$_{ex}$) (assuming 
%similar collisional excitation for both). With only 3 detected lines of {\it c}-C$_3$H$_2$, the constrains on the column density is weak and the derived column density $10^{13}\pscm$ is a lower
%limit corresponding to the emission larger than the telescope beam ( $30''$ equivalent to 3600 AU). We nevertheless can compute a column density
%assuming that the emission of {\it c}-C$_3$H$_2$ comes from the same location as that of {\it c}-C$_3$HD (i.e. 2800 AU). }

The observed main beam temperature $T_{mb}$, for a beam of size $\theta_t$ of a gaussian source of size
$\theta_s$ is
\begin{equation}
\begin{split}
T_{mb} = \eta [J(T_{ex})-J(T_{CMB})] [1-e^{-\tau}] \\
=\frac{\theta_s^2}{\theta_s^2+\theta_t^2} [J(T_{ex})-J(T_{CMB})] [1-e^{-\tau}], 
\end{split}
\end{equation}
where,  $\eta$ is the beam dilution factor which is identical for two molecules since we fix the source size, $J(T)=(h \nu/K) 1/[e^{h\nu/KT}-1]$ is the radiation temperature and $\tau$ is the opacity. For the optically thin conditions, we can 
assume that $\tau \propto N$. Under this assumption, $T_{mb}\propto  [J(T_{ex})-J(T_{CMB})] N $. 

Under optically thin conditions, we can assume that the excitation properties of the two isotopologues are going to be similar which is confirmed by the fact that we find similar values for the excitation temperatures when 
doing the independent analysis. We need more lines, therefore more sensitive observations to see whether temperatures are in fact different. The choice to use the same temperature has also been made in the past for the LTE modeling 
of isotopologues (See for instance \citet{2002ApJ...565..344C}; \citet{2013A&A...560A...3D}). 

Now we can represent the column density (N$_2$) of {\it c}-C$_3$H$_2$ at the excitation temperature (6.3 K) of {\it c}-C$_3$HD as:

\begin{equation}
%\begin{split}
N_2= N_1 [J(T_{ex1})-J(T_{CMB})]/ [J(T_{ex2})-J(T_{CMB})] 
%[(1+(\theta_{t2}/\theta_{s2})^2)/(1+(\theta_{t1}/\theta_{s1})^2)]
%\end{split}
\end{equation}

Where, N$_1$ is the column density of {\it c}-C$_3$H$_2$ at 8.9 K. At frequency 85 GHz, $J(T_{ex1})=7.02$  for $T_{ex1}$ of 8.9 K; $J(T_{ex2})=4.46$ for $T_{ex2}$ of 6.3 K  [$J(T_{CMB})=1.17$ at $T_{CMB}=2.73$ K], we derive a column 
density of $ \rm {N({\it c}}$-$ \rm{C_3H_2)}$=$N_2= 1.78 N_1= 4.7\times10^{13}\pscm$. Thus, we obtain a deuteration fraction $ \rm {N({\it c}}$-$ \rm{C_3HD)}$/$ \rm {N({\it c}}$-$ \rm{C_3H_2)}$=$6.7\times10^{12}/4.7\times10^{13}=14\%$.

%Assuming that uncertainty on N({\it c}-C$_3$H$_2$)$\ll$N({\it c}-C$_3$HD) (since {\it c}-C$_3$H$_2$ lines are brighter), then the relative uncertainty on the deuteration ratio is equal to that of the numerator 
%i.e. N({\it c}-C$_3$HD). The relative uncertainty for N({\it c}-C$_3$HD) is a factor 2 (see Table~\ref{tab.bayes_result}), which translates into the relative uncertainty for the deuteration fraction by a factor of 2 as well i.e. $N({\it c}-C_3HD)/
%N({\it c}-C_3H_2)=15_{-7.5}^{+15}\%$.

If we define the error on column densities of {\it c}-C$_3$H$_2$ and {\it c}-C$_3$HD by errN$_2$ and errN$1$, statistical error on column densities of {\it c}-C$_3$H$_2$ and {\it c}-C$_3$HD by errstatN$_2$ and errstatN$_1$ and relative calibration error by errcal (10\% in both cases), then we can write:
\begin{equation}
\begin{split}
\frac{\mathrm {err} \mathrm {N_2}  } {\mathrm {N_2} }&=\sqrt{(\frac{\mathrm {errstat} \mathrm N_2}  {\mathrm N_2})^2 + (\mathrm {errcal})^2}  \\
                   &=\sqrt{0.02^2+0.1^2}=0.10
\end{split}
\end{equation}

\begin{equation}
\begin{split}
%\frac{\mathrm{err}{N_1}}{N_1} &=\sqrt{(\frac{\mathrm {errstat}N_1}{N_1})^2 + (\mathrm {errcal})^2}  \\
\frac{\mathrm {err} \mathrm {N_1}  } {\mathrm {N_1} }&=\sqrt{(\frac{\mathrm {errstat} \mathrm N_1}  {\mathrm N_1})^2 + (\mathrm {errcal})^2}  \\
                   &=\sqrt{0.1^2+0.1^2}=0.14
\end{split}
\end{equation}

For N({\it c}-C$_3$HD)/N({\it c}-C$_3$H$_2$)=R, %the relative error on deuterium fraction errR/R=$\sqrt{0.13^2+0.51^2}$=0.52 
then 
\begin{equation}
\begin{split}
\mathrm {R_{max}} &=\mathrm {N({\it c}\mbox{-}C_3HD)}_\mathrm{max}/\mathrm {N({\it c}\mbox{-}C_3H_2)}_\mathrm{min}\\
   %         \mathrm {R_{max}} &= \rm {N({\it c}}- \rm{C_3HD)}/ \rm {N({\it c}}- \rm{C_3H_2)}  \\          
           &= 6.7\times(1+0.14)\times10^{12}/4.7\times(1-0.1)\times10^{13}\\
           & =18\%
\end{split}
\end{equation}
       
\begin{equation}
\begin{split}
\mathrm{R_{min}} &=\mathrm {N({\it c}\mbox{-}C_3HD)}_\mathrm{min}/\mathrm {N({\it c}\mbox{-}C_3H_2)}_\mathrm{max}\\
            &= 6.7\times(1-0.14)\times10^{12}/4.7\times(1+0.1)\times10^{13}\\
           & =11\%
\end{split}
\end{equation}

Thus, error on the deuterium fraction can be written as N({\it c}-C$_3$HD)/N({\it c}-C$_3$H$_2$)=$14_{-3}^{+4}\%$. \citet{1988ApJ...326..924B} also measured deuterium fractionation of {\it c}-C$_3$H$_2$ in the order of 5 to 15\% in TMC-1.

\section{1D modelling of the protostellar envelope }

\subsection {The NAUTILUS chemical model with deuteration}

To model the chemistry of {\it c}-C$_3$HD and {\it c}-C$_3$H$_2$ in IRAS 16293, we used the same approach as \citet{2016MNRAS.458.1859M}. For our simulation, we have used the 
2 phase version of the {\tt NAUTILUS} gas-grain chemical model \citep{2016MNRAS.458.1859M, 2015MNRAS.447.4004R, 2015ApJS..217...20W} with deuteration and 
spin chemistry (Majumdar et al. submitted). {\tt NAUTILUS} allows the computation of the chemical composition as a function of time in the gas-phase and
at the surface of interstellar grains. All the equations and the chemical processes included in the model 
are described in detail in \citet{2015MNRAS.447.4004R}. In the current model, several types of chemical reactions are considered in the gas phase by following the kida.uva.2014 
chemical network of \citet{2015ApJS..217...20W} with the recent extension to deuteration and spin chemistry of \citet{2016arXiv161207845M}. 
In the 2 phase version of {\tt NAUTILUS}, there is no differentiation between the species in the mantle and at the surface.  {\tt NAUTILUS} considers 
interaction between gas and grains via four major processes : physisorption of gas phase species onto grain surfaces, diffusion of the accreted species, reaction at the grain surface, and finally by evaporation to the gas phase. 
Our model also considers different types of evaporation processes such as thermal evaporation, evaporation induced by cosmic rays \citep[following][]{1993MNRAS.261...83H}, and chemical desorption as suggested by \citet{2007A&A...467.1103G}. 
We adopt the similar initial elemental abundances reported in  \citet{2011A&A...530A..61H} with a deuterium and fluorine elemental abundance relative to hydrogen of $1.6\times 10^{-5}$ \citep{2006ApJ...647.1106L} and 
$6.68\times 10^{-9}$ \citep{2005ApJ...628..260N} respectively. Here the species are assumed to be initially in an atomic form as in diffuse clouds except for hydrogen
and deuterium, which are initially in H$_2$ and HD forms respectively. For our standard model, we have used a C/O ratio of 0.7  and an ortho-to-para H$_2$ ratio of 3. 

In our current model, all the variables relative to H$_2$ has been modified in terms of nuclear spin states 
ortho-H$_2$ and para-H$_2$. Since, it has been known from long time  that H$_2$, D$_2$, H$_3$$^+$, H$_2$D$^+$, D$_2$H$^+$ and D$_3$$^+$ along with their spin 
isomers are the main species that dictate deuterium fractionation at low temperature \citep[see][]{2014prpl.conf..859C}. Detailed description and benchmarking of our deuterated network with spin 
chemistry is presented in \citet{2016arXiv161207845M}. This network is available on the KIDA\footnote{\url{http://kida.obs.u-bordeaux1.fr/}} website.  
Our network for surface reactions and gas-grain interactions is based on the one from \citet{2007A&A...467.1103G} with added deuteration and spin chemistry from \citet{2016arXiv161207845M}. 

\begin{table*}
\caption{Modeled and observed fractionations for {\it c}-C$_3$H$_2$.}
   \label{tab.deu_frac}
   \hskip -5cm
   \centering
%   \begin{tabular}{lc|c}
   
%   \begin{tabular}{|c|c|c|c}
%       \hline
%       Radius  & 3000 AU  &4000 AU& 7000 AU  \\
%       \hline
%          \emr{\log_{10} [{\it c}-C_3HD]^{a}}   &    -10.15   & -10.07     &  -9.50         \\
%           
%           \emr{\log_{10}[{\it c}-C_3H_2]^{a}}    &    -8.37      &   -8.35      &    -8.33    \\ 
%           
%      %    \emr{ [\chem{D/H}]^{b}}    &    1.3\% &   2\%   &   7\%       \\
%          
%          \emr{[{\it c}-C_3HD/{\it c}-C_3H_2]^{b}}    &    1.7\% &   2\%   &   7\%       \\         
%          
%          
%         \hline
%       \multicolumn{2}{l}{Notes: \emr{^{a}} [X] = N(X)/N(\chem{H_2})}\\
%       \multicolumn{2}{l}{Notes: \emr{^{b}} Deuterium fraction of {\it c}-C$_3$H$_2$}
%
%    \end{tabular}
%\end{table*}

% \begin{tabular}{|c|c|c|c}

\begin{tabular}{|c|c|c|c|c}
       \hline
       Radius  & 3000 AU  &4000 AU& 7000 AU  &   Observed values   \\
        &   &  &  &  ( at 3000 AU)\\

       \hline
          \emr{\log_{10} [{\it c}-C_3HD]^{a}}   &    -10.15   & -10.07     &  -9.50   &  -11.3      \\
           
           \emr{\log_{10}[{\it c}-C_3H_2]^{a}}    &    -8.37      &   -8.35      &    -8.33  &  \emr{\mbox{-}10.45^{c}}  \\ 
           
      %    \emr{ [\chem{D/H}]^{b}}    &    1.3\% &   2\%   &   7\%       \\
          
          \emr{[{\it c}-C_3HD/{\it c}-C_3H_2]^{b}}    &    1.7\% &   2\%   &   7\%  &  14\%   \\

         \hline
       \multicolumn{2}{l}{Notes: \emr{^{a}} [X] = N(X)/N(\chem{H_2})}\\
       \multicolumn{2}{l}{Notes: \emr{^{b}} Deuterium fraction of {\it c}-C$_3$H$_2$}\\
       \multicolumn{5}{l}{Notes: \emr{^{c}} Abundance of {\it c}-C$_3$H$_2$ by considering the same excitation temperature of {\it c}-C$_3$HD (see Section 2.3)}

    \end{tabular}
\end{table*}

\subsection {1D physical structure}
   To follow the deuterium fractionation of {\it c}-C$_3$H$_2$ in IRAS 16293, we have used the same 1D physical structure as in \citet{2008ApJ...674..984A, 2014MNRAS.445.2854W, 2016MNRAS.458.1859M}.  
 This physical structure was based on the nongray radiation hydrodynamic model by \citet{ 2000ApJ...531..350M} to follow the core evolution from pre-stellar core to protostellar core. It starts from a hydrostatic prestellar core with central density 
 $n$(H$_2$) $\sim 3 \times 10^4$ cm$^{-3}$. The core is extended up to $r=4\times10^4$ AU with a total mass of 3.852 $M_{\odot}$, which exceeds the critical mass for gravitational instability.  
 %The outer boundary is fixed at $r=4\times10^4$ AU, so that the total solar mass is 3.852 $M_{\odot}$, which exceeds the critical mass for gravitational instability. 
 Initial temperature for the core is around 7 K at the center and around 8 K at the outer edge. Here, cosmic ray heating, cosmic background radiation, and ambient stellar radiation balance the cooling caused by dust thermal emission. 
 In the model, core stays at its hydrostatic structure for $1\times10^6$ yr to set up the initial molecular conditions for the collapse stage. After $1\times10^6$ yr, the core starts to contract and the contraction is almost isothermal as long as the cooling is efficient. 
 Eventually the compressional heating overwhelms the cooling, which causes rising in temperature in the central region. Contraction then decelerated due to increase in the gas pressure which eventually makes the first hydrostatic core, known the `first core' 
 at the center. When the core center reaches very high density ($10^7$ cm$^{-3}$) and high temperature (2000 K), the hydrostatic core becomes unstable due to H$_2$ dissociation and starts to collapse again. This collapse is referred as the `second collapse'. 
 Within a short period of time, the dissociation degree approaches unity at the center due to rapid increase of central density. Then the second collapse ceases, and the second hydrostatic core, i.e., the protostar, is formed and the infalling envelope around this 
 protostar is known as `protostellar core'.  In the model, the initial prestellar core evolves to the protostellar core in $2.5 \times 10^5$ yr. When protostar is formed, the model further follows the evolution for $9.3 \times 10^4$ yr, during which the protostar 
 grows by mass accretion from the envelope.  
 
% After the onset of contraction, the initial
%prestellar core evolves to the protostellar core in $2.5 \times 10^5$ yr. After the birth of the protostar, the model further follows the evolution for $9.3 \times 10^4$ yr, during which the protostar grows by mass accretion from the envelope.}
% %  In the model, the initial prestellar core evolves to the protostellar core in $2.5 \times 10^5$ yr.
 %Increasing gas pressure decelerates the contraction and eventually makes the first hydrostatic core, known the `first core' at the center. 
% When the core center 
 %becomes as dense as $10^7$ cm$^{-3}$ and as hot as 2000 K, the hydrostatic core becomes unstable due to H$_2$ dissociation and starts to collapse again. This collapse is referred as the `second collapse'. 
%The central density increases rapidly, and within a short period of time the dissociation degree approaches unity at the center. Then the second collapse ceases, and the second hydrostatic core, i.e., the protostar, is formed. The protostar is
%  surrounded by the infalling envelope, which we call the protostellar core. After the onset of contraction, the initial
%prestellar core evolves to the protostellar core in $2.5 \times 10^5$ yr. After the birth of the protostar, the model further follows the evolution for $9.3 \times 10^4$ yr, during which the protostar grows by mass accretion from the envelope.}
 
 \section{Modelling Results and discussions}
Here, we present the predicted abundances of {\it c}-C$_3$HD and {\it c}-C$_3$H$_2$  in the protostellar envelope using the physical and chemical models previously described, and compare with our observations. 
Figure 4 shows the computed abundances of {\it c}-C$_3$HD and {\it c-}C$_3$H$_2$, in the gas-phase and at the surface of the grains in the protostellar envelope as a function of radii, at the end of 
the simulations, i.e. for a protostellar age of $9.3 \times 10^4$ yr. We considered abundances at $9.3 \times 10^4$ yr since the physical structure of the envelope at this age is similar to the one constrained in the envelope of IRAS16293 by 
\citet{2010A&A...519A..65C} from multi-wavelength dust and molecular observations (see \citet{2014MNRAS.445.2854W} for the discussion).

Abundance profiles of {\it c}-C$_3$HD and {\it c}-C$_3$H$_2$ predicted by our model can be divided into three regions. First region is 
defined by radii larger than 200 AU and temperatures below 50 K. In this region, the gas phase abundance of both {\it c}-C$_3$HD and {\it c}-C$_3$H$_2$  decrease toward the centre of the envelope due to high depletion because 
of density increase. In the outer part of the envelope (radii greater than 2000 AU and temperatures below 30 K), gas phase {\it c}-C$_3$HD forms mainly by the dissociative recombination of {\it c}-C$_3$H$_2$D$^+$. 
{\it c}-C$_3$H$_2$D$^+$ is produced mainly from the deuteron transfer from para-H$_2$D$^+$ and DCO$^+$ to {\it c}-C$_3$H$_2$. {\it c}-C$_3$H$_2$D$^+$ is also partly produced by {\it c}-C$_3$H$_2$ + H$_2$DO$^+$ 
and  para-H$_2$ + C$_3$D$^+$ reactions. So in the cold outer envelope, deuteration of {\it c}-C$_3$HD is mainly controlled by para-H$_2$D$^+$ and DCO$^+$. In these region, {\it c}-C$_3$H$_2$ also mainly forms from the dissociative 
recombination reaction of {\it c}-C$_3$H$_3$$^+$. From 2000 AU to 200 AU (where temperature started to increase 
from 30 K to 50 K),  {\it c}-C$_3$HD mainly forms from the neutral-neutral reaction CH + C$_2$HD $\rightarrow$ H + {\it c}-C$_3$HD whereas {\it c}-C$_3$H$_2$ mainly forms from the H + CH$_2$CCH $\rightarrow$ para-H$_2$ + {\it c}-C$_3$H$_2$ reaction.  

Between 200 and 100 AU, the gas phase abundance of both {\it c}-C$_3$HD and {\it c}-C$_3$H$_2$ increases rapidly 
due to the complete evaporation of {\it c}-C$_3$HD and {\it c}-C$_3$H$_2$  from the grain surface. Inside the inner 60 AU, the gas phase abundance of {\it c}-C$_3$HD decreases rapidly due to 
slow deuteration process, i.e. unavailability of any forms of H$_2$D$^+$ to transfer deuteron. But in this region, {\it c}-C$_3$H$_2$ can still survive with a very high abundance ($\sim10^{-9}$) due to its efficient production via the 
reactions H + CH$_2$CCH $\rightarrow$ ortho-H$_2$ + {\it c}-C$_3$H$_2$ and H + CH$_2$CCH $\rightarrow$ para-H$_2$ + {\it c}-C$_3$H$_2$.  

In Table~\ref{tab.deu_frac}, we give our modelled abundances and fractionation ratios at 3000 AU, 4000 AU and 7000 AU. The variation of predicted deuterium fractionation between 3000 AU 
and 4000 AU is very small whereas we observe a large variation at 7000 AU (i.e. in the outer part of the envelope where the temperature is below 30 K). In the outer part of the envelope, H$_3$$^+$ reacts with HD, the major reservoir of D-atoms and D-atom is transferred from HD to H$_2$D$^+$ (ortho/para). As a result, H$_2$D$^+$ becomes very abundant in the outer part and serves as a primary species towards deuterium fractionation. Besides that, other abundant neutrals and important destruction partners of H$_3$$^+$ isotopologues, such as O and CO, deplete from the gas phase (for example because of the freeze out onto dust grains in cold and dense regions). As a consequence, in the outer part, the main formation reaction of {\it c}-C$_3$HD via {\it c}-C$_3$H$_2$D$^+$ + e$^-$ becomes very fast due to efficient deuteron transfer between para-H$_2$D$^+$ and {\it c}-C$_3$H$_2$ to form {\it c}-C$_3$H$_2$D$^+$ which results into an increase in the {\it c}-C$_3$HD abundance steeply beyond 5000 AU. For {\it c}-C$_3$H$_2$, however we have not seen similar behavior due to its efficient destruction via various ion-molecular reactions with H$_3$O$^+$, para-H$_3$$^+$, HCO$^+$, ortho-H$_3$$^+$, and H$^+$. When we are going inside the envelope as compared to outer part, deuteron transfer proceeds via other secondary species (for example at 2000 AU, it is DCO$^+$ which originates from H$_2$D$^+$) which results into lower fractionation. 

%Table~\ref{tab.deu_frac} clearly shows that at 2800 AU, {\it c}-C$_3$HD has the abundance of $5.5\times10^{-11}$ which is less than 
%a factor of 10 deviation from observed abundance of $7\times10^{-12}$. {\bf In current astrochemical models, {\it c}-C$_3$H$_2$ is generally over produced than what has been observed
% (see \citet{2013ChRv..113.8710A} for the review).  Our model produces reasonably high abundance (of the order of $\sim10^{-9}$) of {\it c}-C$_3$H$_2$ as compared to observation ($\sim$$10^{-10}$ by assuming {\it c}-C$_3$H$_2$ emission 
% comes from the same region as {\it c}-C$_3$HD). This translates into lower deuterium fraction from the model as compared to observations. This shows that chemistry of {\it c}-C$_3$H$_2$ in the current astrochemical networks need to 
%be revisited with more laboratory studies and theoretical calculations.}

  The deuterium fractionation predicted by our model at 3000 AU (approximate size of the envelope), is of 1.7\%, i.e. within a factor of 10 than what we obtain from the observations (about 14\%). 
The {\it c}-C$_3$HD modelled abundance is $7\times10^{-11}$ at 3000 AU (see Table~\ref{tab.deu_frac}), which is within a factor of 10 of the observed value ($5\times10^{-12}$). For {\it c}-C$_3$H$_2$, our model over 
predicts the observed one by two orders of magnitude. The discrepancy between the model and the observations could arise from few possible reasons. First, in a highly centrally peaked source like IRAS 16293, the physical structure is of utmost importance to determine the abundance profile of any species. Inhomogeneities in the protostellar envelop could affect the determination of the abundance profile. Although, higher sensitivity observations and higher spectral resolution could help on 
this aspect (since the velocity profile will depend on the physical structure), the real solution would be higher spatial resolution (interferometric) observations. Second, it seems that current astrochemical models over predicts the abundance of {\it c}-C$_3$H$_2$ in different types of sources (see \citet{2013ChRv..113.8710A} for the review; \citet{2016A&A...591L...1S}). Additional chemical studies of the carbon chain productions would have to be done to reproduce the observed abundances.

\section*{Acknowledgements}
Based on observations carried out with the IRAM
30m Telescope. IRAM is supported by INSU/CNRS (France), MPG (Germany) and
IGN (Spain). LM, PG, VW thanks ERC starting grant (3DICE, grant agreement 336474) for funding during this 
work. PG postdoctoral position is funded by the INSU/CNRS. VW acknowledge the CNRS programme PCMI for funding of their research. We would like to thank the anonymous
referee for constructive comments that helped to improve
the manuscript. 
%%%%%%%%%%%%%%%%%%%%%%%%%%%%%%%%%%%%%%%%%%%%%%%%%%



%\bibliographystyle{mnras}


\bibliographystyle{mnras}
\bibliography{C3HD}
\end{document}